\documentclass[letterpaper]{JHEP3}
\usepackage{epsfig}

\newcommand{\be}{\begin{equation}}
\newcommand{\ee}{\end{equation}}
\newcommand{\bear}{\begin{eqnarray}}
\newcommand{\eear}{\end{eqnarray}}
\newcommand{\ba}{\begin{array}}
\newcommand{\ea}{\end{array}}

\newcommand{\CL}{{\cal L}}

\title{The Little Hierarchy in Universal Extra Dimensions
\\ { $\; $ } \\ }

\author{Gustavo Burdman$^\dag$, Alex G. Dias$^{\dag\ddag}$\\
$^\dag$Instituto de F\'{i}sica, Universidade de S\~{a}o Paulo,
\\ R. do Mat\~{a}o 187, S\~{a}o Paulo, SP 05508-900, Brazil \\ \\
$^\ddag$Centro de Ci\^{e}ncias Naturais e Humanas, Universidade Federal do ABC \\
Rua Santa Ad\'{e}lia 166,  
Santo Andr\'{e}, SP 09210-170, Brazil
\\ \\
\email{burdman@if.usp.br, alexdias@if.usp.br} \\ }

\abstract{ \ \

In the standard model in universal extra dimensions (UED) the mass
of the Higgs field is driven to the cutoff of the higher-dimensional
theory. This re-introduces a small hierarchy since the
compactification scale $1/R$ should not be smaller than the weak
scale. In this paper we study possible solutions to this problem by
considering five-dimensional theories where the Higgs field
potential vanishes at tree level due to a global symmetry. We
consider two avenues: a Little Higgs model and a Twin Higgs model.
An obstacle for the embedding of these four-dimensional models in 
five dimensions is that their logarithmic sensitivity to the 
cutoff will result in linear divergences in the 
higher dimensional theory.
We show that, despite the increased cutoff sensitivity of higher
dimensional theories, it is possible to control the Higgs mass in
these two scenarios.  For the 
Little Higgs model studied,  the phenomenology will be significantly
different from the case of the standard model in UED. This is due to the  
fact that the compactification scale approximately coincides with the 
scale where the masses of the new states appear.
For the case of the Twin Higgs model, 
the compactification scale may be 
considerably lower than the scale where the new states appear. If it is 
as low as allowed by current limits, it would be possible to experimentally 
observe the standard model Kaluza-Klein states as well as a new heavy quark. 
On the 
other hand, if the compactification scale is higher, then the 
phenomenology at colliders would coincide with the one for the standard 
model in UED.}

\preprint{hep-ph/0609181}


\begin{document}


\section{Introduction} \setcounter{equation}{0}
\label{intro}

Theories with Universal Extra Dimensions (UED) have attracted
considerable attention recently~\cite{ued}. These theories afford
the possibility that the compactification scale $1/R$ is not far
above the weak scale $M_W$, since the propagation of all fields in
the extra dimensional bulk generates selection rules derived from
momentum conservation. More concretely, after compactification, the
conservation of Kaluza-Klein number in any given vertex involving
Kaluza-Klein (KK) excitations implies that at least two KK modes
must be present in an interaction with a zero mode. As a
consequence, the first KK excitations must be pair produced at
colliders, which lowers the direct search limits with respect to
s-channel production. KK number conservation also means that
electroweak precision constraints can only be affected by one loop
contributions involving KK modes. Although KK-number conservation is
broken by the presence of boundary terms, these typically result in
rather suppressed KK-number violating couplings, leaving the bounds
on $1/R$ still rather low and not much above the weak scale. This
fact has sparked several phenomenological studies about signals for
UED at colliders, where the standard model (SM) in the bulk has been
used as the theory.

However, the SM in UED is not a natural theory, even if $1/R$ is of
the order of the weak scale. This is due to the fact that the bulk
Higgs field is inevitably affected by quadratic divergences that are
only cut off at a scale
$\Lambda_{5D}$ parametrically larger than $1/R$. In order to see this let
us consider a simple model of a scalar field in five dimensions with
the action given by \be S_\Phi = \int d^4x \,dy \,
\left\{\left(D_M\Phi\right)^\dagger D^M\Phi -
V\left(\Phi\right)\right\}~, \label{fiact} \ee with the potential
generically written as \be V\left(\Phi\right) =
M^2\,\Phi^\dagger\Phi + \lambda_5 (\Phi^\dagger\Phi)^2~. \label{pot}
\ee In eqn.(\ref{fiact}) the covariant derivative is $D_M =
\partial_M + ig_5\,A_M$ and the 5D coupling is given by $g_5 =
g\sqrt{2\pi R}$ in terms of the 4D coupling $g$. The theory of
(\ref{fiact}) is non-renormalizable, with a cutoff $\Lambda_{5D}$
defined as the scale where the couplings become strong. For
instance, for the gauge coupling this implies 
\be
N\,\frac{g_5^2}{\ell_5}\,\Lambda_{5D} \simeq 1~, \label{ndag5} 
\ee 
where
$\ell_5 = 24\pi^3$ is the 5D loop suppression factor and $N$ is the size of 
the gauge group.
Eqn.(\ref{ndag5}) defines the cutoff in the usual sense of
na\"{i}ve dimensional analysis (NDA): the scale for which loops are
unsuppressed. With this definition, the interval between the
compactification scale and the cutoff of the theory is 
\be 
\Lambda_{5D} R \simeq \frac{12\pi^2}{g^2\,N}~, \label{lamr} \ee and defines a
maximum value of the Kaluza-Klein (KK) number for which 
the KK modes are weakly coupled.

The central question regarding the scalar theory of
eqn.(\ref{fiact}) is what is the natural value for the mass
parameter $M$. The existence of the potential of eqn.(\ref{pot})
implies the presence of radiative contributions to both $M^2$ and
$\lambda_5$. 
This results in a
contribution to the mass squared schematically given by 
\be
N\,\frac{g_5^2}{\ell_5} \int \frac{d^5k}{k^2} \simeq
\left(\frac{g^2\,N}{36\pi^2}\,\Lambda_{5D} R\right)
 \, \Lambda_{5D}^2~.
 \label{dm2}
\ee 
The first factor in the last expression in eqn.(\ref{dm2}) is of
order one: the loop suppression is canceled at the cutoff
$\Lambda$. 
Thus, radiative corrections naturally give a value 
\be M
\sim \Lambda_{5D} ~, \label{mscalar} 
\ee 
and in the absence of fine
adjustments this should be the typical size of the scalar bulk mass.
If we consider the KK expansion, the KK modes have masses 
\be
m^2_{(n)} = M^2 +\frac{n^2}{R^2}~. \label{kkmass} 
\ee 
Then, if
$\Phi$ were the Higgs doublet in one extra dimension, its zero-mode
would have a mass naturally  at the cutoff $\Lambda_{5D} \gg 1/R$. This
fact is not changed by the presence of fermions or electroweak
symmetry breaking, and it constitutes what we have called the little
hierarchy problem of the SM in universal extra dimensions.

Although not as marked as the hierarchy problem of the SM in four
dimensions, this hierarchy between $1/R$ and $\Lambda$ implies
significant fine tuning in order to keep the Higgs zero-mode light.
A natural solution to this little hierarchy problem  requires a
symmetry in the bulk forbidding the potential for the Higgs field.
Furthermore, this symmetry must be broken so as to generate a
potential leading to a mass of the order of the weak scale for the
scalar. 

In this paper we explore the possibility of
building a model where the Higgs is a (pseudo-) Nambu-Goldstone
boson in UED. Several ideas along this line have been implemented
in 4D in the last few years. Here we will implement a Little Higgs
model~\cite{lhgen,simple1,simple2} and a Twin Higgs model~\cite{twin,twin2,twinlr}
in UED. 

In Little Higgs models 
a global symmetry is spontaneously broken giving rise to
Nambu--Goldstone Bosons (NGBs). Part of the global symmetry
is gauged, so that gauged interactions explicitly break the global
symmetry leading to a radiatively generated potential. The Higgs is
then a pseudo-NGB. The potential is now logarithmically dependent on
the cutoff of the non-linear sigma model. However, when
going to a five-dimensional theory, this tame logarithmic
divergence turns into a linear divergence. We will show that, in spite
of this, it is possible to stabilize the Higgs mass at the weak scale
in Little Higgs theories. As an example we will work in the Simplest
Little Higgs~\cite{simple1,simple2}. The resulting Higgs mass is
somewhat heavier than in the 4D case, $m_h\sim 250~$GeV. This is due
to the effects of the KK modes of the top. The solutions we find 
in this case imply that the compactification scale $1/R$ more or less
coincides with the scale where the zero-modes of the Little Higgs
``partners'' should be. This has important consequences for the
phenomenology of UED models.

We will also implement the Twin Higgs model of Ref.~\cite{twin} in
UED. In the 4D model it is possible to eliminate the cutoff dependence 
coming from the top sector, by introducing extra fermions. We show
this is still the case in the 5D UED theory. Even with this feature,
the Higgs mass is found to be somewhat heavier than in the 4D case, 
in the range $m_h\sim (170-250)~$GeV.
On the other hand, and unlike in the Little Higgs case, it is possible
for the phenomenology to be the virtually the same as for the SM in UED.

In Section~\ref{lh} we will consider a Little Higgs model,
whereas in Section~\ref{th} we study the use of the Twin
Higgs model in UED, both implemented in theories with one 
compact extra dimension. We conclude in Section~\ref{conc}.

\section{A Little Higgs Model in UED}
\label{lh}

In this section we study the possibility of controlling the Higgs
mass by assuming that it is a pseudo-Nambu-Goldstone boson.
The successful attempts in this direction in four dimensional
extensions of the SM require, in addition to having a spontaneously
broken global symmetry, an extension of the gauge group as well as
new fermions. In these scenarios, called Little Higgs
models~\cite{lhgen}, gauge and Yukawa couplings explicitly break the
global symmetry giving the Higgs a mass that is smaller than the 
Nambu-Goldstone boson (NGB) decay constant $f$ by
a loop factor. The new particle content is forced by the global symmetries to
cancel the quadratic divergences in
the Higgs mass, leaving only milder logarithmic divergences. Several
Little Higgs models are available in the literature~\cite{lhgen}.

Here we will implement the model first proposed
in Ref.~\cite{simple1,simple2} in UED. The model has an enlarged gauge
symmetry: $SU(3)_w\times U(1)_X$, which is broken to the SM gauge
group by the vacuum expectation values (VEVs) of two scalar
triplets. The global symmetry under which these fields $\Phi_1$ and
$\Phi_2$ transform is $SU(3)_1\times SU(3)_2$. This is spontaneously
broken to $SU(2)_1\times SU(2)_2$ by the scalar VEVs. Of the 10 NGBs
resulting from this symmetry breaking, 5 will be eaten by the
massive gauge bosons resulting from the gauge symmetry breaking. The
remaining 5 NGBs remain in the spectrum. In the non-linear
description the scalar fields can be parametrized as 
\be \Phi_1 =
e^{i\Pi\frac{f_2}{f_1}}\left( \ba{c}
0 \\
0\\
f_1 \ea \right) \hspace{1cm} 
\Phi_2 = e^{-i\Pi\frac{f_1}{f_2}}\left( \ba{c}
0 \\
0\\
f_2 \ea \right)~, \label{fi1fi2} 
\ee 
with the NGB fields given by
\be 
\Pi =
\frac{1}{f}\left[
\left(\ba{ccc}
\frac{\eta}{2\sqrt{2}} & 0 & 0\\
0 & \frac{\eta}{2\sqrt{2}} & 0\\
0 & 0 & -\frac{\eta}{\sqrt{2}}
\ea \right)
+ \left( \ba{ccc}
0 & 0 & h_1\\
0 & 0 & h_2\\
h_1^* &h_2^*& 0 \ea \right)\right]~. \label{ngbs} 
\ee 
with $\eta$ a singlet
and 
\be h \equiv \left( \ba{c}
h_1\\
h_2 \ea\right)~, \label{hdef} 
\ee 
an $SU(2)_L$ doublet to be
identified with the Higgs field. Although the $SU(3)_w$ generally
breaks explicitly the $\left(SU(3)\right)^2$ 
global symmetry, it respects it at tree level.
The kinetic terms and potential for $\Phi_1$ and $\Phi_2$ are
invariant under both the global and the gauge symmetries. The
explicit breaking induced by the gauge interactions must involve a
power of the operator $\Phi_1^\dagger\Phi_2$, as shown in ~\cite{simple2}. 
In order for the
$SU(3)_w$ to generate this kind of operator it should go through a loop
such as the one shown in Figure~\ref{f1f2}. This diagram radiatively
generates the operator $\left(\Phi_1^\dagger\Phi_2\right)^2$.
However, its cutoff dependence is only logarithmic. 
Thus, only logarithmically divergent loop
contributions to the Higgs mass are induced by the explicit
breaking. The overall scale of these is determined by the vacuum
expectation value breaking the global and gauge symmetries, $f$. The
contribution to the Higgs mass is then of order $f/4\pi$, which is
of the order of the weak scale, if $f\sim O(1)$~TeV. The gauge
symmetry is purposely chosen to explicitly break the global symmetry
only in such a way so as to generate logarithmically divergent scalar
masses. This is at the heart of this as well as other Little Higgs
models.
\FIGURE{
\centering
\epsfig{file=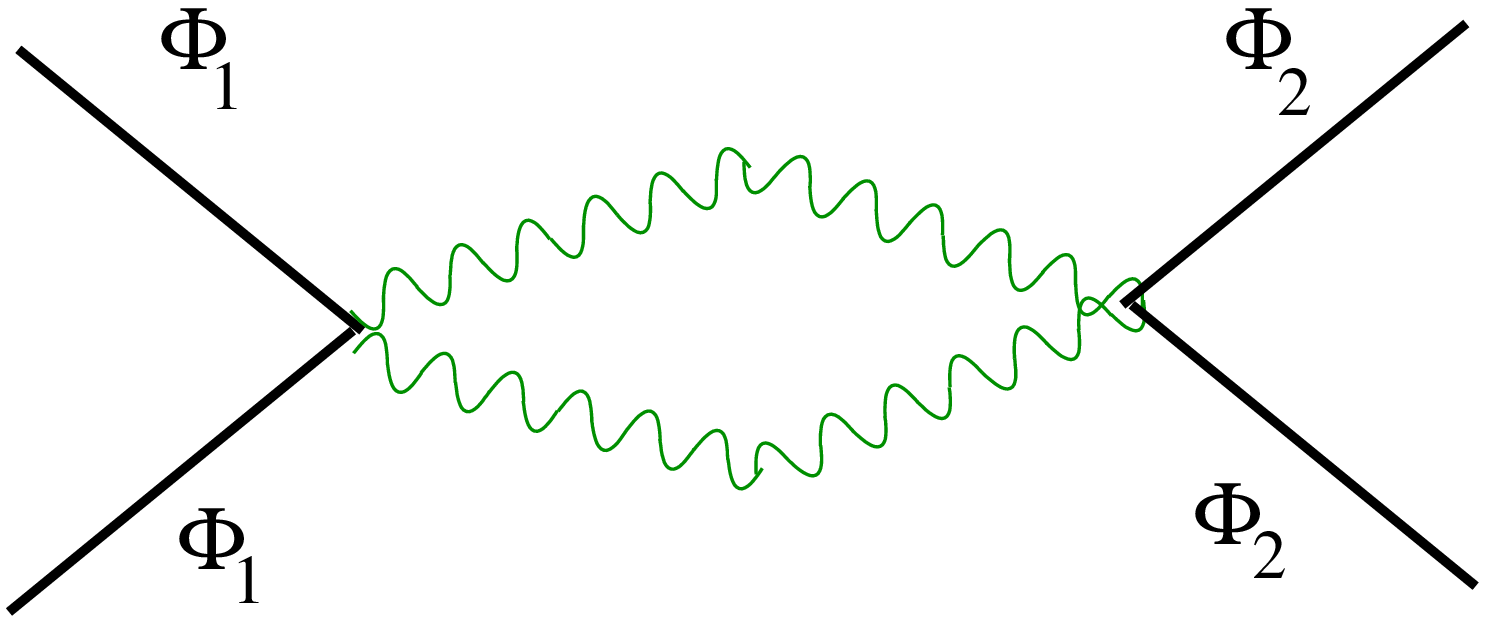,width=6cm,height=3cm,angle=0}
\caption{One-loop contribution from gauge bosons to the operator
$\left|\Phi_1^\dagger\Phi_2\right|^2$.
}
\label{f1f2} 
}

We would like to implement this model in a UED scenario in order to
test if it could be used to solve the Little Hierarchy problem. We
will compute the Coleman-Weinberg potential for the zero mode Higgs
coming not just from the zero mode spectrum of the Little Higgs
model, but also from the KK modes. In order to do this, we
implement the cutoff of the 5D theory as a maximum KK number in the
4D effective (KK) theory. Our NDA estimate of eqn.(\ref{lamr}) tells
us how high the KK number can be before the theory becomes strongly
coupled. For a typical gauge coupling, 
using (\ref{lamr}) results in several dozens of KK modes.
However, the Little Higgs model in UED  defines
a physical cutoff corresponding to that of the non-linear sigma
model underlying the theory. The non-linear description can only be
valid up to 
\be \Lambda \simeq 4\pi f~. \label{nda_nsm} 
\ee 
The
relation between the two cutoffs, and therefore the two scales $f$
and $1/R$, must then be determined. 
We will not consider the case where $f$ is considerably smaller than
$1/R$. Since we still need $f\sim 4\pi v$, this case results in 
values of $1/R$ which would render the extra dimensions 
irrelevant for TeV scale physics. 

Next, we consider $f>R^{-1}$. This includes the case where
both the 5D  and the Little
Higgs theories have approximately the same cutoff,
i.e. $\Lambda_{5D}\simeq \Lambda$. 
Then we have that 
\be f \simeq
\frac{3\pi}{g^2 \,N}\;\frac{1}{R}~. \label{case1} 
\ee 
where $N=N_c = 3$ if we consider the QCD coupling.
In this case, the
Higgs mass is only effectively regulated above the compactification
scale, and therefore can be as large as $1/R$. This would work for
low compactification scales, of the order of the weak scale, but is
not viable when $1/R$ gets to be close to the TeV scale.

\begin{figure}
\centering
\epsfig{file=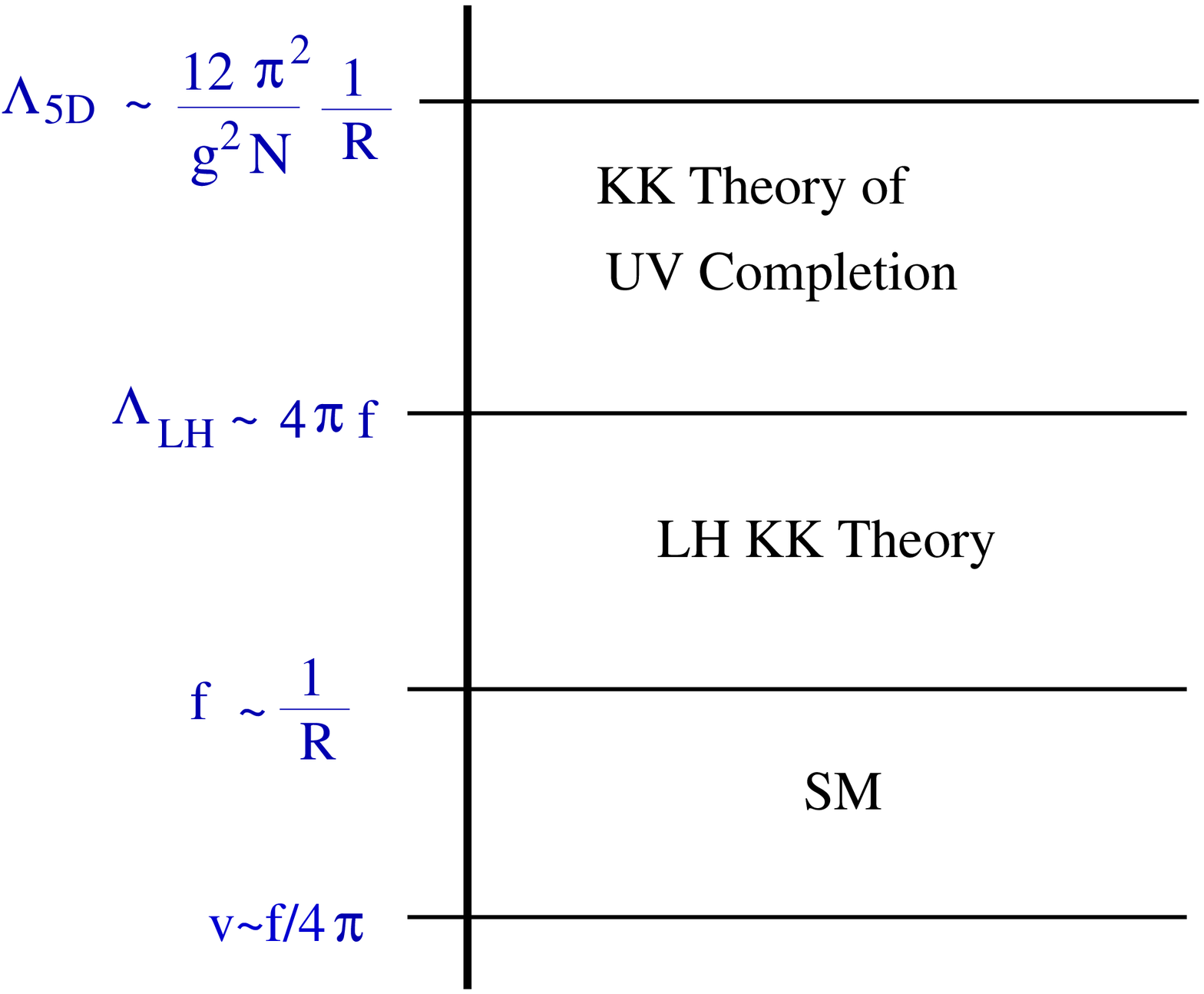,width=8cm,height=7cm,angle=0}
\caption{The case with $f\simeq 1/R$. The contributions to the Higgs
mass come from scales below the little Higgs cutoff, $\Lambda_{LH}\sim
4\pi f$. The KK modes above this scale do not contribute to the Higgs
potential.}
\label{feqrmo}
\end{figure}

Finally, another possibility is for the two scales to be similar 
\be f\simeq
\frac{1}{R}~, \label{fsim1or} 
\ee 
which would yield a Higgs mass
considerably below the compactification scale, allowing us to
accommodate experimental limits on KK modes. In this case, the
cutoff $\Lambda$ of the Little Higgs theory from eqn.(\ref{nda_nsm}) will be
lower than the maximum energy scale for the KK description to be
weakly coupled. 
Above this cutoff of the non-linear sigma model,
there will still be a valid KK mode description, yet the fields
expanded in KK modes will correspond now to the ones appearing in
the ultraviolet completion (UV) of the Little Higgs model. This is 
illustrated in Figure~\ref{feqrmo}. Above the cutoff for the Little
Higgs model, there will be no contributions to the Higgs potential. 
Then, the effective number of summed KK modes is 
$n_m = \Lambda R\simeq 4\pi$.

\subsection{The Mass Spectrum of  the Simplest Little Higgs Model in UED}
\label{lhmass}

We will first review the spectrum of the 4D theory of
Ref.~\cite{simple2}, so we can present its KK expansion later. 
The matter content is composed by fermions
forming $ SU(3)_w\times U(1)_X$ representations, in such a way that
anomaly cancellation does occur when the three families are added
together \cite{331svs} (Model 2 in \cite{simple2}). 
For the quarks we have the following set of
triplets 
\be \Psi_{Q_{1}}=   \left( \ba{c}
d \\
u\\
D \ea \right)_L \sim( {\bf3}^{*},0), \hspace{.5 cm} \Psi_{Q_{2}}=
\left( \ba{c}
s \\
c\\
S \ea \right)_L \sim({\bf3}^{*},0), \hspace{.5 cm} \Psi_{Q_3}=
\left( \ba{c}
t \\
b \\
T \ea \right)_L\sim( {\bf 3}, 1/3), \label{psiq} 
\ee 
and right-handed singlets 
\bear & & u_ R,\,\, c_R,\,\, t_R
\sim({\bf1},2/3), \hspace{.5 cm}
d_R,\,\, s_R,\,\, b_R \sim({\bf1},-1/3), \nonumber\\ \nonumber\\
& & T_R \sim({\bf1},2/3) \hspace{1.8 cm} D_R,\,\, S_R
\sim({\bf1},-1/3). \label{qr} 
\eear 
Here, $D$, $S$, and $T$ are new quarks
which, according to the global symmetry of the little Higgs model,
cancel one loop quadratic divergences for the Higgs mass due $d$,
$s$ and $t$ quarks. The scalar triplets in eqn.~(\ref{fi1fi2})
transform as $ \Phi_1$,$\Phi_2\sim({\bf3},-1/3)$ so that the tree
level Yukawa Lagrangian for the quarks is 
\bear 
\CL_q & = &
\lambda_{d1} \overline{\Psi}_{Q_{1}} \Phi_1^* d_R + \lambda_{d2}
\overline{\Psi}_{Q_{1}} \Phi_2^* D_R   + \lambda_{s1}
\overline{\Psi}_{Q_{2}} \Phi_1^* s_R +
\lambda_{s2} \overline{\Psi}_{Q_{2}} \Phi_2^* S_R   \nonumber\\
& + & \lambda_{t1} \overline{\Psi}_{Q_{3}} \Phi_1 t_R  +
\lambda_{t2} \overline{\Psi}_{Q_{3}} \Phi_2 T_R + h.c. \label{lq}
\eear 
The Yukawa couplings are taken to be diagonal, avoiding
cross terms such as $\overline{\Psi}_{Q_{3}} \Phi_1 T_R $ in
order to simplify the analysis. The quarks $u$, $c$ and $b$ get
their masses through higher-dimensional operators. The dominant 
fermionic contributions to the Higgs mass come from the last two
terms in eqn. (\ref{lq}), corresponding to the top quark and its
partner $T$. Therefore, we disregard all other terms but the
last two  in eqn. (\ref{lq}).

Up to order $v^4/f^4$, the mass squared of  the top quark
and its partner are (see Appendix~A for details)
\bear 
m^2_{t} &=& \left( \lambda_{t1}\lambda_{t2} \right)^2
\frac{f^2 v^2}{M_T^2 } \left[ 1- \frac{1}{3} \left(
\frac{fv}{f_1f_2} \right)^2 + \left( \lambda_{t1}\lambda_{t2}
\right)^2   \left( \frac{f v}{M_T^2 }\right)^2  \right]
\label{mt} \\  \nonumber\\
m^2_{T} &=& M_T^2- m^2_{t}. \label{mT}
\eear

In order to restrict the number of free parameters, we choose  
$\lambda_{t1}$ and $\lambda_{t2}$ so as to minimize $M_T$ for given
values of the 
$f_i$'s, following Ref. \cite{simple2}. 
The
top quark mass then is given by $m_{t}\approx  \lambda_{t1}\lambda_{t2}
 \frac{f v}{M_T }$, and the top quark Yukawa is
 \be
\lambda_{t} = \lambda_{t1}\lambda_{t2}\frac{f}{M_T}
  \label{lt}
 \ee
which fixes $\lambda_{t1}$ and $\lambda_{t2}$ to 

 \be
\lambda_{t1} = \sqrt 2 \lambda_{t} \frac{f_2}{f},
\,\,\,\,\,\,\,\,\,\,\,\,\,\,\,\,\,
 \lambda_{t2} = \sqrt2\lambda_{t}\frac{f_1}{f}
  \label{lmin}~,
 \ee

and the heavy top mass to 

\bear
 M_{T} &=& 2 \lambda_{t}\frac{f_1f_2}{f}.
\label{MTmim}
 \eear

The spectrum's of the zero-mode gauge bosons consists of 
the neutral gauge bosons $Z_1$, $Z_2$
and $U^0$, with masses (up to  order $v^4/f^4$)
\bear
 M^2_{Z_1} &=& m^2_{Z^0}\left[1-
\frac{1}{3}\frac{v^2f^2}{f_1^2f_2^2} +
\frac{m^2_{Z^0}}{M^2_{Z^\prime}} \right]
\label{MZ1} \\
\nonumber\\
M^2_{Z_2} &=& M^2_{Z^\prime} -  M^2_{Z_1} \label{MZ2} \\
\nonumber\\
M^2_{U^0} &=& \frac{1}{2}g^2 f^2 \label{MU0}~, 
\eear
where $m^2_{Z}={g^2v^2}/{2\cos^2\theta_W}$ 
is the tree level squared mass of the SM $Z$.

Also up to  order $v^4/f^4$ the mass squared of the charged diagonal mass
eigenstates $W^{\pm}$ and $W^{\prime\pm}$are
\bear
 M^2_{W} &=& m^2_{W}\left[1-
\frac{1}{3}\frac{v^2f^2}{f_1^2f_2^2} + \frac{m^2_{W}}{M^2_{Ch}}
\right]
\label{MW} \\
\nonumber\\
M^2_{W^{\prime}} &=& M^2_{Ch} -  M^2_{W} \label{MWl} \eear
where $m^2_{W}={g^2v^2}/{2 }$ is the squared mass of the SM $W$
at tree level.

We will now obtain the KK spectrum in the simplest little Higgs 
model.
We consider a theory with one flat extra dimension compactified 
on a $S_1/Z_2$ orbifold.
In order to obtain the desired zero-mode spectrum, we choose the
fermion triplets to satisfy 
$\Psi(x,-y)=-\gamma_5\Psi(x,y)$. 
This results in a KK expansion of the form
\bear
 \Psi(x,y) = \frac{1}{\sqrt{2\pi R}}\psi_{0L}(x) +
 \frac{1}{\sqrt{\pi R}}\sum_{n=1}^\infty
\left[\psi_{nL}(x)\cos(\frac{ny}{R}) +
\hat{\psi}_{nR}(x)\sin(\frac{ny}{R}) \right].
 \label{PsiL}
 \eear
For the singlets, we have $u(x,-y)=\gamma_5 u(x,y)$,
resulting in
\bear {\cal U}(x,y) = \frac{1}{\sqrt{2\pi R}}u_{0R}(x) +
 \frac{1}{\sqrt{\pi R}}\sum_{n=1}^\infty
\left[u_{nR}(x)\cos(\frac{ny}{R}) -
\hat{u}_{nL}(x)\sin(\frac{ny}{R}) \right].
 \label{SingR}
 \eear

The fermion kinetic terms result in the following 4D KK theory: 
\bear \CL_{{4D}} &=&  \int_0^{2\pi R}dy \,\, \overline{\Psi}
i\Gamma^{^M}
\partial_{_M} \Psi ~\nonumber \\\nonumber \\
&=&  \overline{\psi}_{0L} i\gamma^{\mu}\partial_{\mu}{\psi}_{0L} +
\sum_{n=1}^\infty \left[\overline{\psi}_{nL}
i\gamma^{\mu}\partial_{\mu}{\psi}_{nL}+\overline{\hat{\psi}}_{nR}
i\gamma^{\mu}\partial_{\mu}{\hat{\psi}}_{nR}-
\frac{n}{R}\left(\overline{\psi}_{nL}{\hat{\psi}}_{nR}+
\overline{\hat{\psi}}_{nR}{\psi}_{nL}\right) \right]~,\nonumber\\
 \label{L4D}
 \eear
where $\Gamma^{\mu}\equiv \gamma^{\mu}$, $\Gamma^{4}\equiv
i\gamma_{5}$ and $\partial_{4}\equiv \partial_{y}$, will have the
4D Little Higgs model as the zero-mode spectrum. 
Thus, for example, the mass for the top quark KK mode and its
partner have the form
\bear m^2_{n,t} &=& m^2_{t} + \frac{n^2}{R^2 }, \label{mnt}
\\  \nonumber\\
m^2_{n,T} &=& m^2_{T} + \frac{n^2}{R^2 }. \label{mnT}
 \eear

For the gauge bosons, we work in the $A_5=0$ gauge. 
For instance, for the abelian field of the
U(1)$_X$ gauge factor $B_{_M}(x,y)$, 
the 5D kinetic term results in the 4D effective Lagrangian
\bear \CL^{g.b}_{_{4D}} &=&  -\frac{1}{4}\int_0^{2\pi R}dy \,\,
 B^{^{MN}}B_{_{MN}}
\nonumber \\\nonumber \\
&=&  -\frac{1}{4}B_0^{\mu\nu}B_{0\mu\nu}
 + \sum_{n=1}^\infty
\left[ -\frac{1}{4}B_n^{\mu\nu}B_{n\mu\nu}+\frac{1}{2}
\frac{n^2}{R^2}B_n^{\mu}B_{n\mu} \right]
 \label{L4Dbg}
 \eear
and similarly for the non-abelian gauge fields. 

\subsection{The Coleman-Weinberg Potential}
\label{cwzmp}

The one-loop Coleman-Weinberg potential for the zero-mode Higgs
generated by fields having the mass matrix ${\bf
M_i}(v,f)$ is generically given by~\cite{colwein}
\bear 
V &=& \frac{1}{64\pi^2 }\sum_i N_{i}\left [2\Lambda^2 {\bf Tr}
\,({\bf M^\dagger_i M_i}) + {\bf Tr}\left\{ ({\bf M^\dagger_i
M_i})^2\left (\ln  \frac{{\bf M^\dagger_i M_i}}{\Lambda^2}
 -\frac{1}{2}\right) \right\}\right ] \label{veff}~,
  \eear
where $N_{i}$ is the number of degrees of freedom for the case of
bosonic fields, minus the number of degrees of freedom for fermionic fields, and
we have omitted a constant term proportional to the fourth power in
the cutoff.
Due to the Little Higgs global symmetry, the quadratically divergent
term in (\ref{veff}) does not contribute to the Higgs mass.
We write the
relevant part of the potential
generated for the Higgs field, by collecting all
contributions produced by the SM fields and and their partners. 
The contribution of each pair
can be written as
\be
 V_{i} \approx  \frac{N_{i}}{64\pi^2 }\left [ 2 \delta_i \ln \frac{\Lambda^2}{M_i^2}
  + \frac{\delta_i^2}{M_i^4}\left ( \frac{1}{2} +
  \ln \frac{\delta_i}{M_i^4} \right)\right ]~, 
\label{veap} 
\ee
where we have kept terms in $\delta_i$ resulting in 
quadratic or quartic contribution to the 
Higgs potential.
The complete zero-mode contribution to the 
Higgs potential 
will then be $V(h^\dagger h)=\sum_i V_{i}$. 
Keeping up to quartic terms 
we have 
\be
 V(h^\dagger h) = m^2h^\dagger h + \lambda (h^\dagger h)^2~,
 \label{vh} 
\ee
where
\bear
 m^2 &=& - \frac{3}{16\pi^2 } 
 \left [ 2\lambda_t^2\,M_T^2 \,\ln\left(\frac{\Lambda^2}{M_T^2 }\right)
 -\frac{g^2}{4}M_{Z'}^2\, \ln\left(\frac{\Lambda^2}{M_{Z^\prime}^2}\right)
 -\frac{g^2}{2}M_{Ch}^2\,\ln\left(\frac{\Lambda^2}{M_{Ch}^2 }\right)
  \right ]  , \label{m2}
   \eear
and 
\bear
 \lambda = \frac{1}{3 } \frac{f^2}{f_1^2f_2^2 } \vert m^2\vert
 &-& \frac{3}{64\pi^2 v^4}
 \left [4m_t^4\left (\frac{1}{2}+ \ln \frac{m_t^2}{M_T^2 } \right )
 -m_{Z}^4\left ( \frac{1}{2} + \ln \frac{m_{Z}^2}{M_{Z^\prime}^2} \right)
  \right.  \nonumber\\ \nonumber\\
  & & \left. -2m_W^4\left (\frac{1}{2} + \ln \frac{m_W^2}{M_{Ch}^2} \right)
  \right ]~, 
\label{lamb}
 \eear
which agrees with Ref.\cite{simple2}.

As noted in Ref.\cite{simple2}, this potential does not result in
electroweak symmetry breaking for an 
acceptably high scale $f$. This is the result of the mass $m$ in 
eqn.~(\ref{m2}) being too large and negative. This problem was dealt
with in \cite{simple2} by adding the tree-level ``$\mu$'' term 
\bear
   V_{\rm soft} &=& \mu^2 \Phi_1^{\dagger}\Phi_2 + H.c
   \nonumber\\ \nonumber\\
   &\approx& \mu^2 \left [ -2f_1f_2 + \frac{f^2}{f_1f_2}h^\dagger h
   - \frac{1}{12}\frac{f^4}{f_1^3f_2^3}(h^\dagger h)^2
   + \frac{f^2}{2f_1f_2}\eta^2+... \right ]~.
   \label{softb}
 \eear
In addition to lower the Higgs mass, the presence of this term 
explicitly breaks the global $U(1)$ that was keeping the 
$\eta$ massless. With this term present the Higgs mass is now
\bear
   m_H= 2\,v\, \sqrt{\lambda -
\frac{1}{12}\frac{\mu^2f^4}{f_1^3f_2^3} } \,\,
   \label{hm0}~,
 \eear
allowing us to have a light Higgs for reasonably high values of $f$ 
(i.e. $\sim TeV$).

We will now consider the contributions of the KK modes 
to the Coleman-Weinberg potential. As mentioned in Section~\ref{intro}, 
the 5D and the Little Higgs theories are non-renormalizable, and to 
each of them corresponds a cutoff. In the case with both 
cutoffs coinciding, the scale $f$ is significantly higher than 
$R^{-1}$. Then the SM KK theory populates the region 
between $R^{-1}$ and $f$. Above $f$, 
the KK theory now is that for the Little Higgs model. 
In this case, the Little Higgs theory is not efficient regulating
$m_h$, since the large number of KK modes actually contributing to the 
potential result in an unacceptably  large value of $m_h$.

If on the other hand, $f\sim R^{-1}$, then the Little Higgs cutoff
$\Lambda\sim 4\pi f$ appears before the cutoff of the 5D theory. 
Above this cutoff, the KK theory corresponds to the UV completion of
the Little Higgs model, and therefore there will be no contributions to the 
Higgs potential. 
We will then sum the KK 
contributions to the Higgs potential up to a certain KK number
corresponding to the cutoff of the Little Higgs theory, $n_m$. 
Generically, for a given pair, the Coleman-Weinberg potential has the form
\bear
   V^{kk}_{i} &=& 
 \frac{N_i}{64\pi^2}\sum_{n=1}^{n_{_{m}}}\left \{ \delta_i +
\left(\frac{n^4}{R^4}+m^4_{i-} \right)\ln \left (
\frac{n^4}{\Lambda^4R^4} + \frac{n^2M_i^2}{\Lambda^4R^2 } +
\frac{\delta_i}{\Lambda^4 }\right )  \right. \nonumber\\
\nonumber \\
  &+ & \left.   M_i^2\left(\frac{2n^2}{R^2}+ M_i^2-2m^2_{i-}\right ) \ln \left (
\frac{n^2}{\Lambda^2R^2} + \frac{M_i^2}{\Lambda^2 } -
\frac{m^2_{i-}}{\Lambda^2}\right )
\right. \nonumber\\
\nonumber \\
  &+ & \left. 2m^2_{i-}\frac{n^2}{R^2} \ln \left ( \frac{n^2+
m^2_{i-}R^2 }{n^2+M_i^2R^2 - m^2_{i-}R^2}  \right ) \right \}
\nonumber \\ \nonumber \\
 &\simeq & -2\delta_i \,\frac{N_i}{64\pi^2} \left[ n_{_{m}} - 2\ln
\frac{(\Lambda R)^{n_{_{m}}}}{n_{_{m}}!}  \right ].
\label{veapkk}~.
\eear
In the last line we omitted subdominant terms, as well as 
those that do not depend on $v$. The appearance of 
$n_m$, corresponds to the  dependence on the cutoff of the Little
Higgs theory: $\Lambda R = n_m$. Then if we keep only 
the dominant behavior 
with $\Lambda$ we obtain
\bear
    V^{kk}_{i} &\approx & \frac{N_{i}}{32\pi^2} \delta_i
  \left( \Lambda R -
\ln (2\pi \Lambda R)  \right )
 \label{vkkass}~.
\eear
From eqn.~(\ref{vkkass}) we can see that the KK contributions introduce
a linear cutoff dependence in the Higgs potential, to be compared with
the logarithmic dependence for the zero-mode contributions. This is
not surprising. As mentioned in Section~\ref{lh}, the radiatively
generated potential can be seen as coming from the operator 
$\left|\Phi_1^\dagger \Phi_2\right|^2$. In the 4D theory, loop 
diagrams such as the one in Figure~\ref{f1f2}, generated by gauge
boson interactions, give rise to this 
operator with a logarithmic dependence. However, in the 5D theory 
we can see by power counting that they will result in a linear
dependence on $\Lambda$. The relevant interaction in the scalar 
kinetic  term of the 5D theory is 
\be
\left(D_\mu\Phi_i(x,y)\right)^\dagger\,D^\mu\Phi_i(x,y) = 
\cdots + g_5^2\,
A_\mu(x,y)A^\mu(x,y)\,\Phi_i^\dagger(x,y)\Phi_i(x,y)~,
\label{kin5d}
\ee
where $i=1,2$ and $g_5$ is the 5D gauge coupling associated with the 
gauge field $A_M(x,y)$. The one loop contribution to the Higgs mass
resulting from the contribution to $|\Phi_1^\dagger\Phi_2|^2$ will then go like
\be
\frac{g_5^4\,f^2}{(2\pi R)^2}\,(2\pi R)\,\int\frac{d^5k}{(2\pi)^5}\,\frac{1}{k^4} 
\simeq\frac{g^4\,f^2}{6\pi^2}\,\,\left(\Lambda R\right) \label{fddiv}~,
\ee
where we used $g_5=g\sqrt{2\pi R}$ and $g$ is the corresponding 4D gauge
coupling. Thus, a 5D Little Higgs must have linear sensitivity to the 
cutoff $\Lambda$. In what follows we explicitly study how this
feature affects the effectiveness of the Little Higgs mechanism in 
controlling the Higgs mass and having satisfactory electroweak
symmetry breaking. 

The KK modes of the $t$ and $T$  quarks give the following
contributions to the quadratic and quartic terms in the potential

\bear
  m^2_{T_{KK}} =   - \frac{3\lambda_t^2M_{_{T}}^2}{8\pi^2 }
 \left\{  2n_{_{m}}-\ln 2\pi n_{_{m}}
 - \sum_{n=1}^{n_{_{m}}} \left ( \frac{n^2}{M_{_{T}}^2R^2}+1\right )
 \ln \left (1+\frac{M_{_{T}}^2R^2}{n^2}\right )
   \right \}, \label{m2qksh}
   \eear

\bear
 \lambda_{T_{KK}} = - \frac{f^2m^2_{T_{KK}}}{3f_1^2f_2^2}
  - \frac{3\lambda_t^4}{16\pi^2 }
  \left\{2n_{_{m}}- \sum_{n=1}^{n_{_{m}}}
  \left ( \frac{2n^2}{M_{_{T}}^2R^2}+1\right )
   \ln \left (1+\frac{M_{_{T}}^2R^2}{n^2}\right )
 \right\}
 \label{lambqksh}~.
 \eear

The contributions from the KK modes of the gauge bosons of the model
are
\bear
  m^2_{g_{KK}} &= & \frac{3g^2M_{Z^\prime}^2}{64\pi^2 c_W}\left\{
   2n_{_{m}}-\ln 2\pi n_{_{m}}  - \sum_{n=1}^{n_{_{m}}} \left (
\frac{n^2}{M_{Z^\prime}^2R^2}+1\right ) \ln \left
(1+\frac{M_{Z^\prime}^2R^2}{n^2}\right )\right \}
\nonumber\\
\nonumber \\
&+& \frac{3g^2M_{_{Ch}}^2}{32\pi^2 }\left\{
   2n_{_{m}}-\ln 2\pi n_{_{m}}  - \sum_{n=1}^{n_{_{m}}}
\left ( \frac{n^2}{M_{_{Ch}}^2R^2}+1\right ) \ln \left
(1+\frac{M_{_{Ch}}^2R^2}{n^2}\right )\right \}, \label{m2gbksh}~,
\eear
and 
\bear
 \lambda_{g_{KK}} &= &- \frac{f^2m^2_{g_{KK}}}{3f_1^2f_2^2}
  + \frac{3g^4}{256\pi^2c_W  }
  \left\{2n_{_{m}}- \sum_{n=1}^{n_{_{m}}}
  \left ( \frac{2n^2}{M_{Z^\prime}^2R^2}+1\right )
   \ln \left (1+\frac{M_{Z^\prime}^2R^2}{n^2}\right )
 \right\}\nonumber\\
\nonumber \\
&+&\frac{3g^4}{32\pi^2  }
  \left\{2n_{_{m}}- \sum_{n=1}^{n_{_{m}}}
  \left ( \frac{2n^2}{M_{_{Ch}}^2R^2}+1\right )
   \ln \left (1+\frac{M_{_{Ch}}^2R^2}{n^2}\right )
 \right\}
 \label{lambgbksh}~.
 \eear
Then the linear cutoff dependence appears in both the mass squared 
and the quartic coupling.  
In addition to the contributions above, we will also consider the
presence of a tree-level $\mu$ term such as the one described in 
eqn.(\ref{softb}). 
In the following section we present our results for various 
representative values of the parameters and discuss them in detail.

\subsection{Results and Discussion}
\label{numslh}
Putting together all the contributions to the effective potential 
for  the zero-mode Higgs, we will now consider several possible values
for the ratio of VEVs 
\be
k\equiv \frac{f_1}{f_2}
\label{defk}
\ee
as well as for the $\mu$ term defined in eqn.~(\ref{softb}), that
result in electroweak symmetry with $v=246/\sqrt{2}~$GeV. 
For each successful case we obtain the value of the scale $f$ and of
the Higgs mass $m_h$. 
The phenomenological viability of a given solution is determined
mainly by asking $f$ to be high enough for the heavy states not to
have been observed directly, as well as by the requirements from electroweak
precision constraints (EWPC). 
In Table~\ref{stablekk}, we consider the case $f=R^{-1}$, for 
$\mu=0$ as well as two other representative 
values not too different from the weak scale. We also consider
different values of $k$. 
\begin{table}
\centering
\begin{tabular}{||c|c|c|c|c||}\hline
$\mu$ & $k$ & $  f $ & $m_h$ \\
\hline
500  & 1   & 1100  &   246 \\
-    & .5  & 1500  &   249 \\
-    & .25 & 3400  &   246 \\
300  & 1   &  700  &   246 \\
-    & .5 &  950  &   248 \\
-    & .25  &  2100 & 244 \\ 
200  & 1   & 500   &   247\\
-    & .5  & 700  &    249 \\
-    & .25 & 1500  &   245 \\
\hline
\end{tabular}
\caption{
Results for the Higgs mass and the scale $f$, for various values of 
$k=f_1/f_2$ and the soft $\mu$ parameter defined in the text. Note that 
the results for $k\to 1/k$ are virtually identical (see text for explanation). 
Here we used $f=R^{-1}$.}
\label{stablekk}
\end{table}
The results show that it is possible to have a solution with a Higgs mass 
sufficiently light for acceptable values of the scale $f$. This seems
to inevitably require the presence of a $\mu$ term, just like in the 
4D case studied in Ref.\cite{simple2}. The typical value of the Higgs
mass for these solutions is around  $250$~GeV. 
For instance, for $\mu=200~$GeV and $k=0.25$, we obtain a solution
with $m_h\simeq 260$~GeV for $f=1.5~$TeV. This value of $f$ may be at
the edge of what can be accommodated by electroweak 
precision constraints. In order to raise $f$, we must
consider
solutions with larger $\mu$'s, as it can be seen in
Table~\ref{stablekk}. This increases the fine-tuning of the solution (see discussion below). 
Solutions with $\mu = 0$ give values of $f$ that are too low and excluded by experiment.
It should be noticed that, within the approximation made here neglecting the 
contributions from the to the Higgs potential from light fermions,  
the resulting potential 
is symmetric under the exchange $f_1 \leftrightarrow f_2$. Then,  
Table~\ref{stablekk}  can be completed with the entries 
for $k \rightarrow 1/k$, which  are identical to the 
ones shown. 

We then conclude that, for the case in which $f\sim R^{-1}$, is at
least possible to control the Higgs mass with the simplest little
Higgs construction of Ref.\cite{simple2}, at the cost of a somewhat
heavier Higgs. This is due to the linear cutoff sensitivity of 
the 5D theory, compared to the 4D theory of \cite{simple2}, which is 
logarithmic. 

Another distinct case to consider is when $f$ is parametrically larger
than $1/R$. This is of interest, since we can imagine having 
$1/R\simeq v$ and $f$ sufficiently above that, 
still allowed by experimental constraints (direct and
indirect). But, as discussed at the beginning of this section, 
if we take $f>1/R$, this puts the Little Higgs  cutoff $\Lambda$ 
not far from the 
5D cutoff $\Lambda_{5D}$. 
This means that now there would be a larger number of 
KK modes contributing to the Higgs potential. For instance, for
$f=4\pi R^{-1}$ the number of KK modes contributing is larger than
100.  Thus, the linear divergences make it impossible to find 
acceptable solutions. We then conclude that in this context is not
possible to entertain values of $R^{-1}$ as small as $300~$GeV, as
allowed by experiment. This seems to be a very generic feature of 
Little Higgs models in 5D.

The presence of the tree-level $\mu$ term balancing radiative contributions
raises the possibility of fine-tuning in the minimization of the Higgs potential. 
In order to check for this, we consider as a measure of fine tuning the sensitivity
of the Higgs VEV to variations in $\mu$. For this purpose we compute 
\be
\frac{v^2}{\mu^2} \left|\frac{\partial v^2}{\partial \mu^2}\right|~.
\label{ft}
\ee
For  the solutions with $\mu = 200~$GeV, this is always smaller than 10, 
signaling a fine-tuning of better than $10\%$. For example, for $k=0.25$ this resuls in 
a $20\%$ fine-tuning. 
On the other hand, for the solutions with $\mu = 500~$GeV the fine-tuning is already a few percent, 
being about $3\%$ for $k=0.25$ and $6\%$ for $k=1$. Finally, for $\mu = 300$~GeV and $k=0.25$ we find
the fine-tuning to be around $10\%$. Thus, it is possible to obtain solutions with high enough 
values of $f$ (so that they are not experimentally excluded), but with reasonable levels of 
fine-tuning.

Finally, we comment on electroweak precision constraints. 
For the case of interest, $f\sim R^{-1}$, the main 
contributions to oblique corrections are still from the zero-mode new gauge bosons, 
$W^{\prime\pm}$, the non-hermitian $U^0$, the
neutral $Z_2$ and a tree level mixing between
eigenstates $Z$ and $Z^{\prime}$. 
For the $S$ and $T$ parameters we have~\cite{inami}  
\bear
    S & \approx &
  -\frac{1}{4\pi}\frac{M_W^2}{M_{U^0}^2} \left( 5+
  \frac{2}{3\cos^2\theta_W}\right )
 \label{S} \\
\nonumber \\
 T & \approx & \frac{1}{8\alpha}\left( 1- \tan^2\theta_W \right )^2\frac{v^2}{f^2}
  +\frac{1}{4\pi\cos^2\theta_W}\frac{M_W^2}{M_{U^0}^2}
  \left( \frac{3}{2\tan^2\theta_W}\frac{M_{W^\prime}^2}{M_{U^0}^2}-1 \right ).
 \label{T}~,
   \eear
where the first term in eqn.~(\ref{T}) is due the tree-level mixing
$Z-Z^{\prime}$. 
The tree level contribution to the $T$ parameter is
dominant over the one loop corrections. 
For the large values
of the Higgs mass shown in Table~\ref{stablekk}, 
agreement with EWPC is somewhat better due to the positive
contribution to $T$.
For instance, for $\mu=200$~GeV,
$k=0.25$ and $f=1.5$ TeV, it is found $S\approx-0.006$ and $T\approx
0.20$, still allowed at the $1\sigma$ level for 
$m_h\simeq 260$~GeV~\cite{pdg2004}. 
Loop contributions, which dominate $S$, are decoupling. Therefore, keeping 
only the zero-mode contributions gives a good estimate of $S$. 

On the other hand, mixing with the new heavy fermions causes shifts in the 
couplings of gauge bosons with light fermions. Furthermore, four-fermion interactions 
are induced by integrating out the heavy gauge bosons and are tested by LEP2 data. 
A full study of the electroweak precision constraints is outside the scope of this work.
There is currently no study even for the 4D version of the model we used here 
(Model II in \cite{simple2}).
However, a full study of the anomalous model (Model I in \cite{simple2}) 
in the 4D case reveals that the bounds on the scale $f$ are 
stronger than the ones derived from only the oblique corrections~\cite{skiba}. 
If those bounds were to  also apply to the UED version of the non-anomalous model, 
then the allowed solutions would have values of $k$ further from 1 and larger values of the 
tree-level parameter $\mu$, all of which result in larger values of the scale $f$. 
This would push the model to a somewhat more fine-tuned situation, as we discussed above.

The phenomenology of this Little Higgs model in one UED will differ
significantly from the corresponding to just having the SM in the 5D
bulk~\cite{fdphen}. Since $f\simeq 1/R$, the appearance of the KK modes of the SM fields
will be accompanied by the presence of the zero modes of the Little
Higgs model, both for gauge bosons and for fermions. Thus, the impact of these 
new states in phenomenological studies at the LHC cannot be ignored.

\section{A Twin Higgs Model in UED}
\label{th}

We will now study another mechanism to control the
Higgs mass in a UED scenario.  It was recently pointed out in  
that discrete symmetries could be used in addition
to global symmetries, in order to forbid quadratic divergences
from contributing to the Higgs mass~\cite{twin,twin2,twinlr}. 
We will consider a Twin Higgs
model  in one UED to see if this mechanism is effective in controlling
the divergences  arising in extra dimensional theories.

We begin by quickly reviewing the idea behind Twin Higgs models. 
A $Z_2$ discrete symmetry 
is imposed between the SM fields and a ``mirror'' or ``twin'' sector which
transforms under a mirror gauge symmetry. There is also a global
$SU(4)$ symmetry in the Higgs sector. The breaking of the mirror gauge
symmetry down to the SM gauge interactions also breaks $SU(4)\to
SU(3)$. The $Z_2$ symmetry greatly constrains the form of the
contributions to the Higgs potential.
The gauge group is now SU(2)$_A\times$SU(2)$_B$ with the SM
fields transforming under SU(2)$_A$, and their
partners transforming under SU(2)$_B$.
The Higgs doublet corresponds to 4 of the 7 massless degrees of
freedom 
resulting from the $SU(4)\to SU(3)$ spontaneous breaking. 
The Higgs is in the fundamental of the global $SU(4)$ 
\bear
    H=
  \left(
\ba{c}
H_A  \\
H_B \ea \right )
 \label{Htw}
   \eear
where $H_{A,B}$ are doublets of each subgroup SU(2)$_{A,B}$. 
The gauge interactions explicitly break the global symmetry and 
generate a potential of the form
\be
    \Delta V  \propto  \Lambda^2\left (
  g_A^2H_{A}^\dagger H_{A}+g_B^2H_{B}^\dagger
  H_{B}\right )
 \label{vtwh}
   \ee
However, the $Z_2$ symmetry forces $g_A=g_B$, which 
guarantees that the quadratically
divergent contributions in (\ref{vtwh}) are $SU(4)$ symmetric: 
$\Delta V \propto g^2\Lambda^2H^\dagger H$. 
Thus, these quadratic divergences do not contribute to the Higgs
potential. The sensitivity to the cutoff is only logarithmic in the 4D
realization. 
Also in these models is possible to enlarge the global
symmetry in the top sector in a way that renders its contributions
to the Higgs potential finite. 
This is a very interesting possibility since the top
constitutes the dominant contribution to the Higgs mass. It is also 
particularly useful in going to a 5D theory, since the cutoff
sensitivity is linear instead of logarithmic, as we have seen in the
previous section.

First, we review the relevant field content of the Twin Higgs Model
\cite{twin} we are going to extend to the 5D bulk. 
The top quark Yukawa couplings have an approximate 
$SU(6)\times SU(4)\times U(1)$ 
global symmetry, with the 
$(SU(3)_c\times SU(2)\times U(1))_{A,B}$ subgroups gauged. 
The content of this sector is in the following chiral fermions:
$Q_L=({\bf6,\bar
4})$ and $T_R=({\bf\bar 6,1})$, where we showed the 
transformations under $SU(6)$ and $SU(4)$, respectively, and we omit
the U(1) charge. 
Their branchings under [SU(3)$\otimes$SU(2)]$^2$
are
\bear
    Q_L &=& ({\bf 3,2;1,1}) \oplus ({\bf 1,1;3,2})
    \oplus ({\bf 3,1;1,2})\oplus ({\bf 1,2;3,1})
\nonumber \\
 &=&\,\,\,\,\,\,\,\,\,\,q_A\,\,\,\,\,\,\,\,\,\,
 \oplus\,\,\,\,\,\,\,\,\,\, q_B \,\,\,\,\,\,\,\,\,
 \oplus \,\,\,\,\,\,\,\,\,\,\tilde q_A \,\,\,\,\,\,\,\,\,\,\,
  \oplus\,\,\,\,\,\,\,\,\,\, \tilde q_B
    \label{Ql}\\ \nonumber \\
    T_R &=& ({\bf \bar3,1;1,1}) \oplus ({\bf 1,1;\bar3,1})
\nonumber \\
 &=&\,\,\,\,\,\,\,\,\,\,t_A\,\,\,\,\,\,\,\,\,\,
 \oplus\,\,\,\,\,\,\,\,\,\, t_B
    \label{Tr} ~,
   \eear
where once again we omitted the $U(1)$ quantum numbers.

The Yukawa interactions are given by 
\bear
 \CL_q^{twin} & = &  yHQ_LT_R   + h.c
 \nonumber \\ \nonumber \\
  & = &  y(H_At_Aq_A+H_At_B\tilde q_B+H_At_Aq_A + H_Bt_A\tilde q_A+
H_Bt_B q_B)+ h.c \label{lqtw}
 \eear

The exotic quarks $\tilde{q}_A$ and $\tilde{q}_B$ get $Z_2$ symmetric
masses with additional fermions $\tilde{q}_A^c$ and $\tilde{q}_B^c$,
which have the opposite quantum numbers: 
\be
 M(\tilde q_A^c\tilde q_A+\tilde q_B^c\tilde q_B) ~,
\label{vectormass}
\ee
where the mass parameter $M$ is the sole source of 
$SU(4)$ breaking. 

A non-linear realization of the Twin Higgs, has the Higgs in
the broken generators corresponding to the seven Nambu-Goldstone Bosons (NGBs)
of $SU(4)/SU(3)$. Of these, three linear combinations are absorbed by three of the 
$SU(2)_B\times U(1)_B$ gauge bosons. Four degrees of
freedom remain to form the complex Higgs doublet $h$:
\be
 H = e^{i\frac{\Sigma}{f}}\left( \ba{c}
0 \\
0\\
0\\
f \ea \right)   \label{Htw2}
 \ee
with 
\be
 \Sigma=\left( \ba{cccc}
 &               & & h \\
 & O_{3\times 3} & &  \\
 &               & & 0  \\
h^\dagger & 0 & &  0 \ea \right)   \label{ngbtw}
 \ee

The relevant mass terms in the top sector result in~\cite{twin}
\bear
 & & m^2_{tA} \approx
\frac{y^2M^2}{M^2+y^2f^2}v^2,
 \,\,\,\,\,\,\,\,\,\,\,\,
 M^2_{TA} \approx  M^2+y^2f^2 - m^2_{tA},\label{apmTAth}
 \\ \nonumber \\
 & & m^2_{tB} \approx  y^2f^2,
 \,\,\,\,\,\,\,\,\,\,\,\,\,\,\,\,\,\,\,\,\,\,\,\,\,\,\,\,\,\,\,\,\,\,\,
  M^2_{TB} \approx  M^2. \label{apmTBth}
 \eear

The gauge boson spectrum results in (see Appendix~B for details)
\bear
m_{\gamma}^2 &=& 0,\,\,\,\,\,\,\,\,\,\,\,\,\,\nonumber
 \\ \nonumber \\
 m_{Z_A}^2&=&\frac{g^2f^2}{2c_W}\sin^2\frac{v}{f}\nonumber
 \\ \nonumber \\
m_{W_A}^2 &=& \frac{g^2}{2}f^2\sin^2\frac{v}{f}~, \label{magbth}
 \eear
for the $A$ sector, and 
 \bear
M_{\gamma_B}^2 &=& \frac{1}{2} \left
[\frac{M_{Z_A}^2}{\tan^2\frac{v}{f}} +M_B^2 -
\sqrt{\left(\frac{M_{Z_A}^2}{\tan^2\frac{v}{f}}
+M_B^2\right)^2-4M_{W_B}^2M_B^2} \right ],
\\ \nonumber \\
M_{Z_B}^2&=&\frac{1}{2}\left [\frac{M_{Z_A}^2}{\tan^2\frac{v}{f}}
+M_B^2 + \sqrt{\left(\frac{M_{Z_A}^2}{\tan^2\frac{v}{f}}
+M_B^2\right)^2-4M_{W_B}^2M_B^2} \right ],
 \\ \nonumber \\
M_{W_B}^2 &=& \frac{g^2}{2}f^2\cos^2\frac{v}{f}~, \label{mbgbth}
 \eear
for the $B$ sector. Expanding in $v/f$ this spectrum coincides with the 
one obtained in Ref.~\cite{twin}.

\subsection{The Coleman-Weinberg potential for the Twin Higgs Model in UED}
\label{Vth}

We will now compute the effective potential for the zero-mode Higgs, 
starting from the contributions from the top sector. 
The contributions from the zero modes in the top sector result in 
\bear
 m^2_{\rm top} &=&  \frac{3}{8\pi^2 } \frac{y^2M^2}{M^2-y^2f^2}
 \left [ M^2 \ln \frac{M_{TA}^2}{M^2 }
 -y^2f^2 \ln \frac{M_{TA}^2}{m_{tB}^2 }
  \right ], \label{m2qth}
   \eear
for the term proportional to $h^\dagger\,h$, whereas for the quartic 
coupling we obtain
\bear
 \lambda_{\rm top} = &-& \frac{m_q^2}{3f^2}  + \frac{3}{16\pi^2 }y^4M^4
 \left [ \frac{1}{M_{TA}^4}\ln \frac{M_{TA}^2}{m_{tA}^2}
 + \frac{M_{TA}^2}{(M^2-y^2f^2)^3}\ln \frac{M^2}{m_{tB}^2} \right ]
 \nonumber \\\nonumber \\
 &-&\frac{3}{32\pi^2 } y^4M^4\left [\frac{1}{M_{TA}^4} + \frac{4}{(M^2-y^2f^2)^2}  \right ]
 \label{lambqth}
 \eear
in agreement with Ref.~\cite{twin}. 

We will now consider the contributions from KK modes. Just as in the 
case of the Little Higgs in Section~\ref{lh}, the cutoff of the 5D theory is defined 
by strong coupling in the KK theory. However, contributions above the Twin 
Higgs cutoff $\Lambda\simeq 4\pi f$ do not add to the Higgs potential. 
Thus, the sum over KK modes should be cut at $n_{\rm m} \simeq \Lambda R$, with 
$\Lambda$ the Twin Higgs cutoff rather than the 5D cutoff $\Lambda_{5D}$ defined 
in eqn.~(\ref{lamr}). 

In order to obtain the contributions from the KK fermions 
we must be careful in keeping all relevant terms in $v/f$ in eqn.~(\ref{mTBth}).
After collecting the quadratic and quartic terms we obtain
\bear
  m^2_{tKK} =    \frac{3}{8\pi^2 }y^2M^2\sum_{n=1}^{n_{\rm m}}& &\left\{
\frac{1}{M_{_{TA}}^2}
 \left ( \frac{n^2}{R^2}+M_{_{TA}}^2\right )\ln \left (1+\frac{M_{_{TA}}^2R^2}{n^2}\right )
\right. \nonumber \\ \nonumber \\
 & & \left. -\,\,\frac{1}{M^2-y^2f^2} \left (
\frac{n^2}{R^2}+M^2\right )\ln \left (1+\frac{M^2R^2}{n^2}\right )
\right. \nonumber \\ \nonumber \\
 & &\left. + \,\, \frac{1}{M^2-y^2f^2} \left ( \frac{n^2}{R^2}+y^2f^2\right ) \ln \left
(1+\frac{y^2f^2R^2}{n^2}\right )
   \right \}, \label{m2qkth}
   \eear
and 
\bear
 \lambda_{tKK} = - \frac{m^2_{tKK}}{3f^2} & +& \frac{3}{16\pi^2 }y^4M^4
  \sum_{n=1}^{n_{\rm m}}
  \left ( \frac{2n^2}{R^2}+M_{_{TA}}^2\right )
  \left\{ \frac{1}{M_{TA}^6}\ln \left (1+\frac{M_{_{TA}}^2R^2}{n^2}\right )
 \right. \nonumber \\ \nonumber \\
 & & \left. +\,\, \frac{1}{(M^2-y^2f^2)^3}\left [\ln \left (1+\frac{M^2R^2}{n^2}\right
 )- \ln \left
(1+\frac{y^2f^2R^2}{n^2}\right )\right ]
 \right\}
 \nonumber \\ \nonumber \\
& &- \,\,\frac{3}{8\pi^2 }n_{\rm m} y^4M^4\left [\frac{1}{M_{TA}^4}
+ \frac{1}{(M^2-y^2f^2)^2}  \right ]~. 
 \label{lambqkth}
 \eear
As expected from the fact that the Yukawa couplings are $SU(4)$
symmetric, the zero-mode contributions from eqns.~(\ref{m2qth}) and 
(\ref{lambqth}) are finite and regulated by $M$. 
This is also the case for the KK contributions in eqns.~(\ref{m2qkth})
and (\ref{lambqkth}). Although one could imagine that summing over the
KK modes might re-introduce cutoff sensitivity through the dependence
on $n_{\rm m}\simeq\Lambda R$, this is not the case. This can be seen, for
instance, by taking the limit of large $n$ in eqns.~(\ref{m2qkth})
and  (\ref{lambqkth}) and seeing that all terms involving $n_{\rm m}$ 
cancel. The sum over the KK modes does not introduce any 
dependence on the cutoff $\Lambda$ in the top sector. We then conclude that the $SU(4)$
symmetric Yukawa couplings are still efficient in regulating the 
contributions from the top sector KK modes. This will be very
important in obtaining suitable solutions for electroweak symmetry
breaking with a relatively light Higgs.
 
The gauge boson contributions are not regulated and therefore do lead
to cutoff sensitivity. The zero-mode contributions to the mass
parameter are given by 
\bear
 m^2_{_{g}} &=& \frac{3}{64\pi^2 }\left( 3g^2M_{W_B}^2 \ln\frac{\Lambda^2}{M_{W_B}^2}
 + g^{\prime 2}M_{Z_B}^2 \ln \frac{\Lambda^2}{M_{Z_B}^2}\right), \label{m2gbth}
   \eear
in agreement with Ref.~\cite{twin}. The contributions to the quartic
coupling are
\bear
 \lambda_{_{g}} = -\frac{5}{6} \frac{m^2_{_{WZ}}}{f^2}  &+ &\frac{3}{256\pi^2f^2 }
 \left [ 3g^2M_{W_B}^2\left ( 1- 2\ln\frac{\Lambda^2}{M_{W}^2} \right)
 +2g^{\prime 2}M_{Z_B}^2   \right ]
 \label{lgbth}
 \eear

The contributions from the gauge boson KK modes read
\bear
  m^2_{gKK} &= & \frac{3}{64\pi^2 }\left (3g^2M_{W_B}^2+g^{\prime 2}M_{Z_B}^2 \right)
  \left ( 2n_{\rm m}-\ln 2\pi n_{\rm m} \right) \nonumber \\ \nonumber \\
  & -& \frac{3}{64\pi^2 }\sum_{n=1}^{n_{\rm m}}\left\{
3g^2\left ( \frac{n^2}{R^2}+M_{W_B}^2\right )\ln \left
(1+\frac{M_{W_B}^2R^2}{n^2}\right )
\right. \nonumber \\ \nonumber \\
 & &\left. + \,\, g^{\prime 2}\left (
\frac{n^2}{R^2}+M_{Z_B}^2\right )\ln \left
(1+\frac{M_{Z_B}^2R^2}{n^2}\right )
   \right \}, \label{m2gbkth}
   \eear
for the quadratic term in the potential. For the quartic coupling they
are given by
\bear
 & &\lambda_{gKK} = - \frac{m^2_{gKK}}{3f^2} -\frac{3}{64\pi^2f^2 }
   \left (3g^2M_{W_B}^2+g^{\prime 2}M_{Z_B}^2 \right)
   \left ( n_{\rm m}-\ln 2\pi n_{\rm m} \right)
 \nonumber \\ \nonumber \\
 & +&\frac{3}{128\pi^2 f^2}\sum_{n=1}^{n_{\rm m}}
 \left\{ 3g^2M_{W_B}^2 \ln \left (1+\frac{M_{W_B}^2R^2}{n^2}\right
 )+\frac{1}{2}g^{\prime 4}f^2\ln \left
(1+\frac{M_{Z_B}^2R^2}{n^2}\right )
 \right\}
 \label{lgbkth}
 \eear
Unlike the contributions from the top sector KK modes, the gauge boson
KK modes do result in linear divergences. This appears as a linear 
dependence on $n_{\rm m} \simeq\Lambda R$.
However, this will not result in a heavy Higgs, since the gauge boson
contributions tend to lower the Higgs mass. 

\subsection{Results and Discussion}
We combine all the contributions to the Coleman-Weinberg potential
obtained in the previous section to look for solutions with
electroweak symmetry breaking and a light Higgs. We consider two
cases: $f=1/R$ and $f=3/R$. 

In the first case, with $f=1/R$, the cutoff of the 5D theory is
well above the Twin Higgs model cutoff of $\Lambda_{TH}\simeq 4\pi
f$. Then, just as in the Little Higgs model of Section~\ref{lh}, we
only sum KK mode contributions up to $n_{\rm m}\simeq \Lambda_{TH}
R=4\pi$.
\begin{table}
\centering
\begin{tabular}{||c|c|c|c||}\hline
$M_B$ & $M$ & $ f $ & $m_h$ \\
(TeV)   & (TeV) & (GeV) & (GeV) \\
\hline
1 & 1.5 & 850 &   230 \\
1  & 3     & 2000 &   560 \\
2 & 2 & 1200  &  270 \\
3  & 1.5     & 1300 &   280 \\
5 & 2 & 850  &  170 \\
\hline
\end{tabular}
\caption{Solutions for $f=1/R$ for the Twin Higgs Model in
one UED. The number of KK modes are summed up to $n_{\rm m} = 4\pi$.}
\label{th1}
\end{table}
In Table~\ref{th1} we show some representative solutions for this case
in which the scale $f$ is adequately large and the Higgs mass not too
heavy. Solutions with larger values of $M$ would tend to have larger 
values of $m_h$ since $M$ regulates the contributions from the top
sector to the Higgs mass. On the other hand, solutions with larger
values of $M_B$ result in a smaller $m_h$, since $M_B$ enhances the 
gauge boson contributions, which make the Higgs lighter. 
The additional fields in the Twin Higgs model have no or very little
couplings
with the SM fields, with the main exception being the Higgs.   
Thus, the phenomenology of this scenario is very similar to the one
for the SM in one UED. The exception is the presence of a heavy
singlet quark plus an $SU(2)_L$ doublet, colorless twin-quark. 
The zero modes of these states appear only at the scale $M$, which
could be not too
far above the compactification scale $R^{-1}$. The twin gauge bosons
are, in principle decoupled from SM fields. The exotic quarks
$\tilde{q}_A$ and $\tilde{q}_B$ 
induce kinetic mixing between the SM photon and the twin photon. 
However, this does not affect collider phenomenology in any significant way.

The motivation for the second case we study is to find a scenario 
where $f$ and $1/R$ are sufficiently separated so as to make it
possible to have an energy interval above the current experimental
limits of $1/R>300$~GeV, where the theory is just the SM in one UED,
without any additional states and with a rather low $1/R$. As an example we
consider $f=3/R$. In Table~\ref{th2} we show representative examples
in this scenario.
\begin{table}
\centering
\begin{tabular}{||c|c|c|c||}\hline
$M_B$ & $M$ & $ f $ & $m_h$ \\
(TeV)   & (TeV) & (GeV) & (GeV) \\
\hline
1 & 2 & 1300 & 500 \\
2  & 1     & 950 &  300 \\
3 & 2 & 1600  &  320 \\
4  & 2     & 1700 &   270 \\
\hline
\end{tabular}
\caption{Solutions for $f=3/R$ for the Twin Higgs Model in
one UED. The number of KK modes are summed up to $n_{\rm m} = 12\pi$.}
\label{th2}
\end{table}
In this case, the sum over the KK modes must be extended up to
$n_{\rm m}=\Lambda_{TH}R\simeq 12\pi$. 
As before, the lighter Higgs masses are obtained for larger values of
$M_B$.  For instance, the last entry in Table~\ref{th2} is 
$M_B=4~$TeV, $M=2~$TeV and $f=1.7~$TeV, which corresponds to 
$R^{-1}\simeq 570~$GeV. This is in agreement with current
experimental bounds, both on $R^{-1}$ as well as on the new 
states associated with the scale $f$. 
Then in these kind of scenarios, it is possible to stabilize the 
Higgs mass in what it would appear  -- at the Tevatron and perhaps
even  at the early LHC-- as the SM in one universal extra 
dimension. However, the LHC should eventually discover the
extended fermion sector. 
Their zero modes  would have masses of order $M$, and their KK
modes would start not far above, typically at $M^2 + 1/R^2$. 
Thus, the discovery at the LHC of the Twin Higgs states would always 
be accompanied by the discovery of a few of the corresponding KK
modes. 

We finally note that, unlike in the 4D case, 
a $\mu$ term was not necessary in order to obtain acceptably light Higgs masses. 
The use of a $\mu$ term would result in even smaller values of $m_h$ relative 
to what is shown in Table~\ref{th2}.

\section{Conclusions}
\label{conc}
Theories with universal extra dimensions, 
in which the entire SM field content propagates in the 5D bulk, 
suffer from a little hierarchy,
as discussed in Section~\ref{intro}. 
We have seen that it is possible to control the Higgs mass if the
Higgs is a pseudo-Goldstone boson living in the 5D bulk. 
This is so despite of the fact that the divergences in these 5D
theories are linear in the cutoff, whereas in 4D they are
logarithmic. 
We considered two scenarios to solve this little hierarchy problem:
a Little Higgs model and a Twin Higgs model.

As an example of the Little Higgs case we studied the Simplest Little
Higgs model of Ref.~\cite{simple2} in one universal extra dimension.
The cutoff dependence, which is logarithmic in four-dimensional Little
Higgs models, is linear in 5D models and tends to make the Higgs
heavier. 
Despite this difficulty, we have found solutions  with relatively low
Higgs masses $m_h\simeq (250-260)~$GeV, for values of the symmetry
breaking scale $f$ that are high enough to satisfy direct as well as
indirect constraints. These solutions though, correspond to the case
in which $f\sim 1/R$. Then, the phenomenology of this solution to the
Little Hierarchy in UED theories is markedly different than the one
corresponding to the SM in UED~\cite{fdphen}, since the appearance of the first KK
modes of the SM fields is always accompanied by the zero-modes of the 
new states in the Little Higgs spectrum, such as the new gauge bosons
and fermions. 

We also considered a Twin Higgs model in one universal extra
dimension. Just as in the Little Higgs case, we are able to stabilize
$m_h$ despite the linear divergences. We do so without the need of a 
$\mu$ term, which was necessary in the Little Higgs case, as well as 
in the Twin Higgs in 4D. 
Furthermore, in the Twin Higgs case we are also
able to find solutions where the scales $f$ and $1/R$ are separated,
in addition to solutions with $f\sim1/R$. Thus, the use of the Twin
Higgs mechanism allows the UED scenario with only the SM fields to be realized,  
with $1/R$ not far above the weak scale and $f$ at or somewhat above
the TeV scale, as it can be seen in Table~\ref{th2}. In both Twin
Higgs scenarios, $f\sim 1/R$ and $f>1/R$, the phenomenology is very
similar to that of the SM in UED, due to the fact that the new Twin
states appearing at the scale $f$ and above tend to have no or very
little interactions with the SM fields. The exception is the presence
of heavy quarks at the scale $M$. Then, generically, the Twin Higgs
mechanism stabilizes the Higgs mass in the extra 
dimensional theory without major changes in the phenomenology.  
This is particularly true for $f\simeq 1/R$, since in this case 
$M$ --and with it the new heavy quark state-- are likely to be 
beyond the reach of the LHC. 

We then conclude that if the KK modes corresponding to the SM in UED
are observed without additional states, the mechanism for stabilizing  
the Higgs mass is probably the Twin Higgs, but certainly not a Little 
Higgs. 

We notice that 6D UED theories, of great interest due to a variety of 
theoretical~\cite{6d1} and phenomenological~\cite{6d2} issues, 
have a stronger cutoff dependence. 
From an argument analogous to the one leading to eqn.~(\ref{fddiv}), we see that
the 6D embedding of a theory with the required global symmetries to protect the 
Higgs mass, would result in a quadratic cutoff dependence. 
This will always be the case as long as the 
explicit symmetry breaking leads to a logarithmic 
cutoff dependence in the corresponding 4D theory. 
Repeating 
the argument of eqn.~(\ref{fddiv}) for the simplest Little Higgs case, now the 
one loop contribution to $|\Phi_1^\dagger\Phi_2|^2$ results, in the 6D UED theory, in a 
contribution to the Higgs mass of the form
\be
\frac{g^4\,f^2}{16\pi}\,\,\left(\Lambda R\right)^2 \label{sddiv}~,
\ee 
which depends quadratically on the cutoff of the Little Higgs theory. 
In the KK picture, this will appear as a quadratic dependence on the the maximum 
number of KK modes to be summed. Then in principle, stabilizing the Higgs mass in a 6D
theory appears to be a rather difficult task. It would probably be necessary to have an 
additional global symmetry in the top sector, such as in the Twin Higgs model presented in
Section~\ref{th}, in order to protect the Higgs mass from the large contributions 
of this sector, which if quadratic in the cutoff would probably make the Higgs unacceptably 
heavy.

\bigskip

{\bf Acknowledgments:}
 We would like to thank Z.~Chacko for many helpful
discussions and comments. 
The authors acknowledge the support of the State of S\~{a}o Paulo
Research Foundation (FAPESP). G.B. also thanks the Brazilian  National Counsel
for Technological and Scientific Development (CNPq) for partial support.

\section*{Appendix A}
\renewcommand{\theequation}{A.\arabic{equation}}
\setcounter{equation}{0}
In this Appendix we derive the expressions for the mass matrices both of 
fermions and gauge bosons used in Section~\ref{lhmass}. 

In order to obtain the mass matrices, we expand the fields in
eqn.~(\ref{fi1fi2}) 
taking $\langle h \rangle= (0 \,\, v)^T$.
Expanding the exponentials in the VEVs results in
\be \langle \Phi_1 \rangle = f_1 \left( \ba{c}
is_1 \\
0\\
c_1 \ea \right) \hspace{1cm} \langle\Phi_2 \rangle = f_2 \left(
\ba{c}
-is_2\\
0\\
c_2 \ea \right)~,
 \label{vevfi1fi2}
  \ee
where we have defined 
\bear
s_1\equiv\sin\left(\frac{f_2v}{f_1f}\right) &&
~ ~ ~ c_1\equiv\cos\left(\frac{f_2v}{f_1f}\right) \nonumber \\
s_2\equiv\sin\left(\frac{f_1v}{f_2f}\right) &&
~ ~ ~ c_2\equiv\cos\left(\frac{f_1v}{f_2f}\right).
\label{trigdef}  
\eear
The mass eigenvalues for a given particle type 
can be written as
\bear 
m^2_{\pm} & = & \frac{1}{2} \left[
M(f)^2 \pm \sqrt{M(f)^4 - 4\delta(f,v)} \right]
\nonumber\\  \nonumber\\
& \approx & \frac{1}{2} \left[ M(f)^2 \pm M(f)^2 \right] \mp
\frac{\delta(f,v)}{M(f)^2} \left[ 1 + \frac{\delta(f,v)}{M(f)^4}
\right] . \label{meig} 
\eear 
In eqn.~(\ref{meig}), $M(f)^2$ is the dominant part of 
the heavy partner mass, mostly associated with the
eigenvalue $m^2_{+}$, and all dependence of the Higgs VEV $v$
appears only in $\delta(f,v)$. 
In what follows we will compute 
$M(f)^2$ and $\delta(f,v)$ for fermions and gauge bosons.

We first consider the fermion spectrum, and in particular the 
top and its partner as it is the dominant fermion contribution. 
Their quadratic  masses are obtained squaring
the mass matrix of the bi-linears in $t$ and $T$ 
\bear 
\CL_{t,T} & =
&  \overline{\Psi}_{Q_{3}} \left( \ba{cc} \lambda_{t1} \langle
\Phi_1 \rangle & \lambda_{t2} \langle \Phi_2 \rangle \ea \right )
\left( \ba{c}
t_R  \\
T_R \ea
\right ) + h.c. \nonumber\\
& = &  \overline{\Psi}_{Q_{3}} {\cal M}_{Q_3}^{\dagger} \left(
\ba{c}
t_R  \\
T_R \ea \right ) + h.c.~, \label{bq} 
\eear 
which gives 
\be 
{\cal M}_{Q_3} {\cal M}_{Q_3}^{\dagger} = \left( \ba{cc}
\lambda_{t1}^2  \langle \Phi_1 \rangle^{\dagger} \langle \Phi_1
\rangle &
\lambda_{t1}\lambda_{t2} \langle \Phi_1 \rangle^{\dagger} \langle \Phi_2 \rangle \\
\lambda_{t1}\lambda_{t2} \langle \Phi_2 \rangle^{\dagger} \langle
\Phi_1 \rangle & \lambda_{t2}^2 \langle \Phi_2 \rangle^{\dagger}
\langle \Phi_2 \rangle \ea \right ). \label{mmtT} 
\ee 
This results in 
\bear 
\delta_t &=& \lambda_{t1}^2\lambda_{t2}^2 f_1^2f_2^2\sin^2
\left( \frac{fv}{f_1f_2} \right)
\label{dt}\\  \nonumber\\
M^2_{ T} &=& \lambda_{t1}^2f_1^2+ \lambda_{t2}^2f_2^2. 
\label{MT}
\eear 
Then, up to order $v^4/f^4$, the mass squared of  the top quark
and its partner are
\bear 
m^2_{t} &=& \left( \lambda_{t1}\lambda_{t2} \right)^2
\frac{f^2 v^2}{M_T^2 } \left[ 1- \frac{1}{3} \left(
\frac{fv}{f_1f_2} \right)^2 + \left( \lambda_{t1}\lambda_{t2}
\right)^2   \left( \frac{f v}{M_T^2 }\right)^2  \right]
\label{Amt} \\  \nonumber\\
m^2_{T} &=& M_T^2- m^2_{t}. \label{AmT}
\eear

We now consider the zero-mode gauge boson spectrum of the model.
From the nine gauge bosons of the $SU(3)_w\times U(1)_X$ sector
eight linear combinations will absorb  eight of the twelve
degrees of freedom in the scalar triplets. The bi-linears of the
gauge bosons come from the following part of the Lagrangian

\bear \CL_{g.b}=\sum_{i=1,2} \left| \left(g W^a_\mu T^a
-\frac{1}{3}g_XB_\mu \right) \langle \Phi_i \rangle \right|^2
\label{lgb} \eear

Among the  neutral gauge bosons there will be two 
massive states, $Z_1$ and $Z_2$, resulting from combinations of $W_3$,
$W_8$ and $B$, two states from combinations of $W_5$ and $W_5$ that we
call $U^0$, and the photon.

\bear
  \CL^{^{mass}}_{neut.} & = & \left( \ba{cccc} W_3 & W_8 & B & W_5 \ea \right )
 \frac{{\cal M}^2_{neu}}{2}\left(
\ba{c}
W_3  \\
W_8 \\
B \\
W_5 \ea \right ) + \frac{g^2}{4}f^2 W_4^\mu W_{4\mu} \label{bneu}
  \eear
After electroweak symmetry breaking, the neutral gauge boson squared
mass matrix is given by
\be
  {\cal M}^2_{neu}  = \frac{g^2}{2}\left( \ba{cccc}
a & \frac{1}{\sqrt3}a & -\frac{2b}{3}a & -u \\
\frac{1}{\sqrt3}a & -a+\frac{4}{3}f^2 & -\frac{2b}{3\sqrt3}(3a-2f^2) & 
\frac{1}{\sqrt3}u \\
-\frac{2b}{3}a & -\frac{2b}{3\sqrt3}(3a-2f^2) & \frac{4b^2}{9}f^2 & 
\frac{4b}{3}u \\
-u & \frac{1}{\sqrt3}u & \frac{4b}{3}u & f^2 \\
 \ea \right ) \label{mmz1z2}
  \ee
with the definitions

\bear
 a &=& f_1^2s^2_1  + f_2^2s^2_2
\label{pa}\\
\nonumber\\
b^2 &=& \frac{3t^2}{3-t^2}
\label{bdef} \\ \nonumber \\
u &=& f_1^2s_1c_1  - f_2^2s_2c_2. \label{pu}
 \eear

Diagonalization of eqn.~(\ref{mmz1z2}) results in 

 \bear
 \delta_z &=& g^4\, \frac{(1+t^2)}{3-t^2} 
f_1^2f_2^2 (s_1c_2  + s_2c_1)^2
\label{dz}\\
\nonumber\\
M^2_{Z^\prime} &=& \frac{2}{3-t^2}\,g^2f^2, \label{MZ}
 \eear
with $t=g^\prime/g = \tan(\theta_W)$.

Therefore, up to  order $v^4/f^4$, the masses of  $Z_1$, $Z_2$
and $U^0$ are
\bear
 M^2_{Z_1} &=& m^2_{Z^0}\left[1-
\frac{1}{3}\frac{v^2f^2}{f_1^2f_2^2} +
\frac{m^2_{Z^0}}{M^2_{Z^\prime}} \right]
\label{AMZ1} \\
\nonumber\\
M^2_{Z_2} &=& M^2_{Z^\prime} -  M^2_{Z_1} \label{AMZ2} \\
\nonumber\\
M^2_{U^0} &=& \frac{1}{2}g^2 f^2 \label{AMU0}~, 
\eear
where $m^2_{Z}={g^2v^2}/{2\cos^2\theta_W}$ 
is the tree level squared mass of the SM $Z$.

For the charged gauge bosons we have four states which are linear
combinations of the symmetry eigenstates as
$W_I^{\pm}=1/\sqrt2(W_1\mp iW_2)$ and $W_{II}^{\pm}=1/\sqrt2(W_6\pm
iW_7)$. 
Schematically we have
\bear
  \CL^{^{mass}}_{ch} & = & \left( \ba{cc} W_I^{+} & W_{II}^{+} \ea \right )
 {\cal M}^2_{W}\left(
\ba{c}
W_{I}^{-}  \\
W_{II}^{-} \ea \right ) \label{bch}
  \eear
with the squared mass matrix given by 
\be
  {\cal M}^2_{W}  = \frac{g^2}{2}\left( \ba{cc} a \,\,\,\,\,&
-iu \\
iu\,\,\,\,\, & f^2-a \ea \right ) \label{mmwW}
  \ee
This results in the parameters of the charged sector spectrum 
\bear
 \delta_{W} &=& \frac{g^4}{4}f_1^2f_2^2 \left(s_1c_2 + s_2c_1
 \right)^2
\label{dw}\\
\nonumber\\
M^2_{Ch} &=& \frac{1}{2}g^2f^2, \label{MCH}~. 
\eear
Then, up to order $v^4/f^4$ the mass squared of the diagonal mass
eigenstates $W^{\pm}$ and $W^{\prime\pm}$ which are mixing of
$W_{I}^{\pm}$ and $W_{II}^{\pm}$ are
\bear
 M^2_{W} &=& m^2_{W}\left[1-
\frac{1}{3}\frac{v^2f^2}{f_1^2f_2^2} + \frac{m^2_{W}}{M^2_{Ch}}
\right]
\label{AMW} \\
\nonumber\\
M^2_{W^{\prime}} &=& M^2_{Ch} -  M^2_{W} \label{AMWl} \eear
where $m^2_{W}={g^2v^2}/{2 }$ is the squared mass of the SM $W$
at tree level.

\section*{Appendix B}
\renewcommand{\theequation}{B.\arabic{equation}}
\setcounter{equation}{0}
Here we show in some detail the calculation of the mass matrices 
for fermions and gauge bosons, used in the text to compute 
the Coleman-Weinberg potential in the Twin Higgs model.
The VEV of the scalar doublet, $\langle h \rangle=v$,  leads to

\be
 \langle H\rangle = \left( \ba{c}
if\sin \frac{v}{f}  \\
0\\
0\\
f\cos \frac{v}{f} \ea \right) = \left( \ba{c}
\langle H \rangle_A  \\
 \\
\langle H \rangle_B \ea \right) \label{Hvev}
 \ee
With this VEV structure we can also find the mass eigenvalues in a
closed form, which will be particularly useful in computing the 
Coleman-Weinberg potential in the 5D UED theory.
This non-linear description is valid up to the cutoff $\Lambda\simeq 4\pi f$.

The mass matrix in the quark sector is
\bear
  \CL_q^{mass} & = & \left( \ba{cccc} t_A & t_B & \tilde q_A^c & \tilde q_B^c \ea \right )
 \left( \ba{cccc}
y\langle H \rangle_A^\dagger & 0 & y\langle H \rangle_B^\dagger & 0 \\
0 & y\langle H \rangle_B^\dagger & 0 & y\langle H \rangle_A^\dagger \\
0 & 0 & M & 0 \\
0 & 0 & 0 & M \\
 \ea \right )\left(
\ba{c}
q_A  \\
q_B \\
\tilde q_A \\
\tilde q_B \ea \right ) +h.c\label{lqth}
 \eear
which results in the eigenvalues 
\bear
 m^2_{tA}&=&\frac{1}{2}\left[
M^2+y^2f^2-\sqrt{(M^2+y^2f^2)^2-4\delta_1} \right ]
\nonumber \\ \nonumber \\
 m^2_{TA}&=&\frac{1}{2}\left[
M^2+y^2f^2+\sqrt{(M^2+y^2f^2)^2-4\delta_1} \right ]
\nonumber \\ \nonumber \\
 m^2_{tB}&=&\frac{1}{2}\left[
M^2+y^2f^2-\sqrt{(M^2+y^2f^2)^2-4\delta_2} \right ]
\nonumber \\
 m^2_{TB}&=&\frac{1}{2}\left[ M^2+y^2f^2+\sqrt{(M^2+y^2f^2)^2-4\delta_2}
\right ]
\label{mTBth}~,
 \eear
with
\bear
 \delta_1=y^2f^2M^2\sin^2\frac{v}{f}
 \label{d1th}\\ \nonumber \\
 \delta_2=y^2f^2M^2\cos^2\frac{v}{f}
 \label{d2th}
 \eear

The gauge boson mass matrix is obtained from the interactions with the 
pNGBs
\bear
 \CL_{g.b}=\sum_{j=A,B} \left| \left(g W^a_{j\mu} \tau_j^a
-\frac{g^\prime}{2} B_{j\mu} \right) \langle H_j \rangle
\right|^2
 \eear
where $\tau_j^a$ are the usual SU(2) generators.
We also consider an explicit mass term $M_B$ for the abelian gauge field 
$B_{B\mu}$, which results in a 
mass for the twin photon and 
breaks the twin symmetry softly. This mass could be generated
dynamically at scales above the cutoff $\Lambda$.
The mass eigenvalues for the $A$ sector can be written as
\bear
m_{\gamma}^2 &=& 0,\,\,\,\,\,\,\,\,\,\,\,\,\,\nonumber
 \\ \nonumber \\
 m_{Z_A}^2&=&\frac{g^2f^2}{2c_W}\sin^2\frac{v}{f}\nonumber
 \\ \nonumber \\
m_{W_A}^2 &=& \frac{g^2}{2}f^2\sin^2\frac{v}{f} \label{Bmagbth}
 \eear
and for the sector $B$
 \bear
M_{\gamma_B}^2 &=& \frac{1}{2} \left
[\frac{M_{Z_A}^2}{\tan^2\frac{v}{f}} +M_B^2 -
\sqrt{\left(\frac{M_{Z_A}^2}{\tan^2\frac{v}{f}}
+M_B^2\right)^2-4M_{W_B}^2M_B^2} \right ],
\\ \nonumber \\
M_{Z_B}^2&=&\frac{1}{2}\left [\frac{M_{Z_A}^2}{\tan^2\frac{v}{f}}
+M_B^2 + \sqrt{\left(\frac{M_{Z_A}^2}{\tan^2\frac{v}{f}}
+M_B^2\right)^2-4M_{W_B}^2M_B^2} \right ],
 \\ \nonumber \\
M_{W_B}^2 &=& \frac{g^2}{2}f^2\cos^2\frac{v}{f}~. \label{Bmbgbth}
 \eear

\newpage

\end{document}